# Crystallographic Orientation Dependent Reactive Ion Etch in Single Crystal Diamond

Ling Xie[†], Tony X. Zhou[†], Rainer J. Stöhr, and Amir Yacoby*

Dr. L. Xie

Center for Nanoscale Systems

Harvard University

Cambridge, MA 02138, USA

T. X. Zhou, Prof. A. Yacoby

Department of Physics, Harvard University

17 Oxford Street

Cambridge, MA 02138, USA

[†] These authors contributed equally to this work.
E-mail: zhou01@g.harvard.edu

Dr. Rainer J. Stöhr

3rd Institute of Physics, Research Center SCoPE and IQST,

University of Stuttgart

70569 Stuttgart, Germany

*(Abstract)*

Sculpturing desired shapes in single crystal diamond is ever more crucial in the realization of complex devices for nanophotonics, quantum computing, and quantum optics. The crystallographic orientation dependent wet etch of single crystalline silicon in potassium hydroxide (KOH) allows a range of shapes formed and has significant impacts on MEMS (microelectromechanical systems), AFM (atomic force microscopy), and microfluidics. Here, a crystal direction dependent dry etching principle in an inductively-coupled plasma reactive ion etcher is presented, which allows to selectively reveal desired crystal planes in monocrystalline diamond by controlling the etching conditions. The principle is demonstrated when the kinetic energy of incident ions on diamond surfaces is reduced below a certain threshold leading to anisotropic etching and faceting along specific crystal planes. Using the principle, monolithic diamond nanopillars for magnetometry using nitrogen vacancy centers are fabricated. In these nanopillars, a half-tapering angle up to 21° is achieved, the highest angle reported, which leads to a high photon efficiency and high mechanical strength of the nanopillar. These results represent the first demonstration of crystallographic orientation dependent reactive ion etch principle, which opens a new window for shaping specific nanostructures which is at the heart of nanotechnology. It is believed that this principle will prove to be valuable for structuring and patterning of other single crystal materials as well.

*(Introduction)*

The ability to transform single crystalline materials into desired shapes is vital in nanotechnology. In micro and nano scale fabrication, controlling etch direction is essential to achieve specific shapes in single crystal materials required by device applications. Developing new etching techniques and processes are therefore critical for successful realization of complex devices. Focused ion beam for sculpting specially shaped individual elements[1] and ion beam milling for controlled angle etch[2] are two examples of sculpting techniques that are largely insensitive to crystalline directions. Taking advantage of the anisotropic nature of monolithic materials, crystal direction dependent wet etching techniques have been developed and are most well-known for etching Si in KOH. In a KOH solution, the kinetics of chemical reactions vary on Si {100}, {110}, and {111} planes, leading to a crystallographic dependent etch[3–9]. This wet etch recipe can process large amount of samples in parallel and is one of the most important assets in modern MEMS technology[10]. Having such similar technique in a dry etch process would be desirable as a tool for more advanced MEMS fabrication. In addition, it would be particularly desirable in nanoelectromechanical systems (NEMS) because wet chemical processes can be difficult to control precisely, especially in case of delicate nanoscale devices. Here, we demonstrate that crystal direction dependent etch can be achieved as a dry process on single crystal diamond.

Diamond is a metastable allotrope of carbon, where the carbon atoms are arranged in a variation of a face-centered cubic crystal structure called a diamond lattice. It has broad applications in science and technology due to its mechanical strength[11], chemical inertness, thermal properties[12], and wide-band optical transparency[13]. Diamond further serves as a host material for a variety of atomic defects, some of which show interesting quantum-mechanical spin and optical properties[14–17]. The presence of such atomic color centers has given diamond an important role in quantum computing[18–20,20,21], magnetometry[17,22–26], and photonics[13,27,28]. For many such applications, optimizing the diamond structure in relation to the color center on the micro and nanoscales and along particular crystal directions is important[27,29–32].

Developing new etching techniques and processes is a critical step for successfully fabricating devices in diamond. Dry etching processes were developed to make anti-reflection coating[33,34], solid immersion lens (SIL)[1], nano-cavities[28,35,36], nano-pillars[27,29,31], nanobeams[28,32,37–41], and microdisk[34,36]. Faceting in diamond was observed in cleaving[11,42], pure chemical etching in high temperature furnaces filled with $O_2$ gas [43,44], and reactive ion etching at high ICP power and high substrate temperature[36]. Though {111} crystal planes were observed at zero substrate power, its mechanism remained unexplored. Thus, crystallographic orientation dependent dry etching has not yet been controlled. In this work, we demonstrate that anisotropic etching along multiple crystal directions in diamond is achieved by controlling the oxygen plasma conditions in an inductively-coupled reactive ion etcher (ICP-RIE). Further, we present the underling etching principle to shed light into diamond crystal direction dependent etching mechanism. Using this principle, Si-KOH etch is resembled on diamond as an encouraging sign that the principle can be applied to other single crystal materials.

*(Why & how crystal direction dependent etch principle can be achieved with RIE - Fig. 1)*

In RIE, etching mechanisms include chemical reactions on exposed surfaces that form volatile byproducts and physical ion bombardment to enhance etch rate and directionality. Major factors controlling etch dynamics include (i) reactive ion flux impinging exposed surfaces, which mainly depends on the concentration of reactive ions in plasma, (ii) the kinetic energy of ions that arrive onto exposed surfaces, which is determined by the negative DC bias between plasma and substrates without considering collisions in the cathode charge region[45], and (iii) the energy barrier for chemical reactions taking place, which is determined by substrate materials and can be anisotropic in certain single crystals. Under a constant reactive ion flux, the etching process is dominated by either the ion's kinetic energy or the energy barrier for chemical reactions. Only when the ion energy is closely tuned to this energy barrier and the chemical reaction limits the etching process, does a high etch selectivity along crystallographic directions emerge. This principle is demonstrated in this work.

**Figure 1** shows a schematic illustration of the formation of a V-shaped groove in a condition of crystal direction dependent etching. Starting with a rectangular etch window defined on a <100> major face oriented crystal substrate, if the etch rate in <hkl> direction is slower than that in the vertical direction <100>, tapered {hkl} sidewalls will develop and grow until a V-shape is fully formed.

*(Experimental conditions)*

Throughout this work, monocrystalline (100) diamond is used. the diamond surface is polished and strain relieved to achieve an rms roughness of less than 1 nm[46]. Prior to processing, diamond substrates are cleaned in a boiling mixture consisting of equal parts sulfuric acid, nitric acid, and perchloric acid to remove organic contaminations and to oxygen terminate the surface. To define an etch mask using electron beam lithography, a layer of flowable oxide (FOX) is spin coated on the sample using a 10 nm thick titanium layer as adhesion promoter. On each sample, one group of rectangular etch windows is aligned with its edges parallel to the <110> direction, while another group of windows is aligned parallel to the <100> direction. The crystal orientation of the samples is independently verified in nitrogen vacancy (NV) center magnetometry experiments revealing <111> crystal axis[47]. Etching experiments are conducted in Plasma-Therm Versaline ICP – RIE system using 900 W ICP power, 40 sccm $O_2$ flow rate, 10 mTorr pressure, and 10 °C substrate temperature by varying the substrate power from 0 W to 120 W. An etch depth of 2 – 3 µm is achieved for each sample by adjusting the etch duration for each given substrate power. For etch rates at various substrate powers and more details on fabrication, please see the supporting information.

*(Crystal direction dependent etch in <110> oriented etching windows - Fig. 2)*

As an example for the crystal orientation dependent etch, **Figure 2** shows several forms, such as a truncated square pyramid in Figure 2 a-c, a V-shaped groove and a truncated rectangular pyramid in Figure 2d. These forms are from the etching masks aligned to <110> direction and etching at 5 W substrate power for 70 min. The faceted sidewalls belong to the {332} family and have an angle of 25° with respect to {110} vertical planes. Flat etched surfaces and fine straight intersection lines are observed at corners and between sidewalls and the bottom surface, as shown by the high magnification image in Figure 2e. The visible polishing marks of ~1nm rms roughness on the top surface are due to the initial polishing of the diamond and are not a result of the etching process. In contrast, the roughness of the etched sidewalls is not resolved indicating an rms roughness of far less than 1 nm. For the square ring shown in Figure 2a, additional facets appeared around the outside corners which are symmetric with respect to the <100> diagonals. Their intersections with the bottom (100) surface are close to <740> direction with an average angle of 60.4° to <100>. These corner facets are the result of different ion fluxes and diverted ion trajectories at the corners of the etch window. As shown in Figure 2d, for steady ion bombardment along the long sides of the rectangular features, except at the end corners, only one facet was developed.

*(Different etching morphologies in <100> oriented etching windows – Fig. 3)*

On the same sample, when the etch masks are aligned parallel to the <100> direction, the resulting etch profiles are dramatically different. Additional surfaces at inner corners emerged as shown in **Figure 3** a-d. These surfaces have an orientation very close to {111} but are not flat. Similarly, the etched sidewalls parallel to <100> are curved and their intersections with the corner surfaces form arc lines. These results imply that the faceting did not fully develop in <100> oriented windows at 5 W substrate power. However, when the substrate power was decreased further to 0 W, the faceting along {111} planes at corners and {100} vertical sidewalls did appear as shown in Figure 3e, as indicated by the straight intersection lines and smooth flat etched surfaces. This crystal faceting at zero substrate power was also observed under etching conditions of 3000 W ICP power and 250 °C substrate temperature as reported in[36].

*(Desired crystal planes can be revealed by varying substrate powers- Table 1)*

For etch masks with edges parallel to <110>, the angles between the faceted surfaces etched at different substrate powers and {110} vertical planes were measured with SEM and the respected Miller indices are assigned accordingly (shown in **Table 1**). These faceted planes have relatively low indices, intersect with <110> direction, and rotate around <110> axis from {111} family to {331} family as the substrate power increased from zero to 40 W, as illustrated in the inset of Table 1. These results imply that desired crystal planes can be revealed by varying the substrate power.

**Table 1.** Angles between faceted surfaces etched at different bias-power for etching windows aligned parallel to <110>

| RF substrate power (W) | DC bias (V) [i] | Angle Θ | Crystal planes |
|---|---|---|---|
| 0 | 9 | 35 | {111} |
| 5 | 15 | 25 | {332} |
| 10 - 20 | 22 - 34 | 19 | {221} |
| 30 - 40 | 47 - 60 | 12 | {331} |

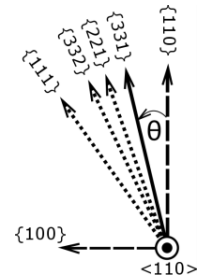

(i) Please see supporting information for full measurement values of DC bias voltage vs. RF substrate power.

*(Crystal direction dependent etch disappears when the ion energy is greater than a threshold)*

In contrast to the low power regime where selectivity is observed, at high substrate powers the etching anisotropy diminished. Figure 3 f-g shows the images of a sample etched at 80 W substrate power with windows aligned to <110> and <100>, respectively. The faceting disappears and cavities formed at corners due to heavier ion bombardment and the etching morphologies become identical despite these two differently oriented etching masks. Assuming that ions have no collisions after they move into the cathode charge region above the substrates[45], the potential energy of ions in the plasma will be entirely transferred to the kinetic energy when ions arrive at the substrate surfaces. Therefore, we find kinetic energy ≤ 60 eV (40W substrate power) to be the threshold to observe crystallographic etching in diamond. It demonstrates that diamond crystallographic etching follows the dry etching principle presented above.

*(Application in photonics and color center research- Fig. 4)*

A direct application of this crystallographic orientation dependent etching technique is the fabrication of monolithic nanopillars with large tapering angles. Such waveguiding structures have been shown to increase the photon collection efficiency of color centers in diamond due to a combination of optical wave guiding and adiabatic changes of the effective local refractive index[31]. As shown in **Figure 4** a-d, tapering angles (half apex angle) of up to 21° are achieved using the presented technique. The taper angle as a function of the substrate power is measured using SEM and shown in Figure 4e. The angle varies linearly with the substrate power when it is ≥ 40 W and is discrete at the lower power region. The switching effect indicates that etching mechanism enters the surface chemical reaction limited regime where kinetic energy of ions and crystalline bonding energy play major roles. This is consistent with the appearance of {331} facets at 30 - 40 W and {221} at 10 – 20 W shown in Table 1.

To verify the waveguide properties of the different nanopillar geometries, the saturated fluorescence intensity of a single NV center implanted 10 nm below the pillars top surface is measured. A group of 10-20 nanopillars containing single NV centers is studied at each taper (half apex angle) angle. All nanopillars shown here have a top diameter of 350 nm and a length of 1.5 μm. Their average saturation count rate and standard deviation are shown in Figure 4f. The results are found to be consistent with previous reports that larger taper angle yields higher photon collection efficiency[31]. With the technique presented here, larger taper angle compared to previous studies can be achieved. Beyond higher fluorescence collection efficiency, the larger taper angle of nanopillars strongly enhances the mechanical strength of the structure. This is particularly important in NV center-based scanning probe applications where the diamond nanopillar is scanned in contact over the sample surface.

In summary, the crystal direction dependent reactive ion etching principle is presented for selectively revealing crystal planes in monocrystalline diamond by varying etching conditions in an ICP-RIE system. The physical intuition is to adjust the reactive ion energy to become comparable with the energy barrier for chemical reactions to take place on crystal planes, which leads to crystal direction dependent etch rates. As a direct application of this technique, we demonstrate the fabrication of monolithic diamond nanopillars with tapering angles of up to 21°, which yielding high photon collection efficiency from single NV centers and high mechanical strength. We believe that the etching principle presented here is applicable to other single crystal materials that hold crystallographic anisotropy and in other types of dry etching systems, such as reactive ion beam etch. This method will enable forming a wide range of shapes in different single crystal materials for a broad variety of applications.

**Experimental Section**

Diamond Substrate Preparation

The electronic grade diamonds (4x4x0.5mm$^3$) provided by Element Six are cut and polished by Delaware Diamond Knifes. Cleaning the diamonds in a boiling mixture of equal parts of sulfuric, nitric, and perchloric acid is required to remove contamination and impurities. Subsequently, one surface of each diamond substrate is strain relieved using an Ar/Cl and O$_2$ RIE process. By doing so, a few µm of the top diamond surface is removed. This top layer is known to have a large concentration of defects and dislocations accumulated during the polishing process. The smoothness of the diamond surface also benefits from this strain relief process as shown in a previous report[1]. The strain relief etch parameters are presented in **Table S1**.

Diamond Fabrication

The diamond is mounted onto a Si carrier chip (1x1cm$^2$) with strain relieved surface facing up using crystal bond for easy handling. This can be done using a hot plate at 150-180°C to melt crystal bond. To promote the adhesion between e-beam resist and substrate, 10nm of Ti is evaporated onto the substrate. Three layers of flowable oxide (FOX-16, Dow Corning) are spin coated with each spun at 3000 RPM for 45s and baked at 100 °C for 10min. After baking, the FOX layer is about 1µm thick. The FOX layer is then directly exposed with e-beam lithography at 100 keV energy and 54µC per cm$^2$ dosage. The exposed FOX layer is developed in 25 wt. % TMAH for 30s followed by a DI water rinse and IPA cleaning. This forms the etch mask for the RIE process. First, an Ar/Cl recipe is used to remove the 10nm Ti layer in the regions not covered by FOX. This exposes the bare diamond surface for the O$_2$ etch process described in Table S1 under supporting information. After the RIE process, the substrate is dipped in HF to remove residual Ti and FOX.

**Supporting Information**

Supporting Information is available from the Wiley Online Library or from the author.

**Author Contributions**

L. Xie and T.X. Zhou contributed equally to this work.

**Acknowledgments**

Sample fabrication was performed at the Center for Nanoscale Systems (CNS), a member of the National Nanotechnology Coordinated Infrastructure (NNCI), which is supported by the National Science Foundation under NSF award no. ECCS - 1541959. CNS is part of Harvard University. We thank Mathew Markham and Element Six (UK) for providing diamond samples. We also thank Dr. Ronald Walsworth and Prof. Matthew Turner for annealing diamonds, and Dr. Mike Burek, Prof.

Marko Loncar, and Prof. Frans Spaepen for fruitful discussions. This work is supported by the Gordon and Betty Moore Foundation's EPiQS Initiative through Grant GBMF4531. Prof. A. Yacoby is also partly supported by the ARO grant W911NF-17-1-0023. Dr. L. Xie acknowledges the support in part by National Science Foundation NNCI program.

References:

[1] M. Jamali, I. Gerhardt, M. Rezai, K. Frenner, H. Fedder, J. Wrachtrup, *Rev. Sci. Instrum.* **2014**, *85*, 123703.

[2] W. Däschner, M. Larsson, S. H. Lee, *Appl. Opt.* **1995**, *34*, 2534.

[3] D. B. Lee, *J. Appl. Phys.* **1969**, *40*, 4569.

[4] K. E. Bean, K. E. Bean, *IEEE Trans. Electron Devices* **1978**, *25*, 1185.

[5] H. Seidel, L. Csepregi, A. Heuberger, H. Baumgärtel, *J. Electrochem. Soc.* **1990**, *137*, 3612.

[6] H. G. G. Philipsen, J. J. Kelly, *J. Phys. Chem. B* **2005**, *109*, 17245.

[7] H. G. G. Philipsen, N. J. Smeenk, H. Ligthart, J. J. Kelly, *Electrochem. Solid-State Lett.* **2006**, *9*, C118.

[8] A. Gupta, B. S. Aldinger, M. F. Faggin, M. A. Hines, *J. Chem. Phys.* **2010**, *133*, 044710.

[9] K. P. Rola, I. Zubel, *Microsyst. Technol.* **2013**, *19*, 635.

[10] M. Gad-el-Hak, Ed. , *MEMS: Design and Fabrication*, CRC/Taylor & Francis, Boca Raton, FL, **2006**.

[11] J. E. Field, *Rep. Prog. Phys.* **2012**, *75*, 126505.

[12] L. Wei, P. K. Kuo, R. L. Thomas, T. R. Anthony, W. F. Banholzer, *Phys. Rev. Lett.* **1993**, *70*, 3764.

[13] I. Aharonovich, A. D. Greentree, S. Prawer, *Nat. Photonics* **2011**, *5*, 397.

[14] T. Iwasaki, F. Ishibashi, Y. Miyamoto, Y. Doi, S. Kobayashi, T. Miyazaki, K. Tahara, K. D. Jahnke, L. J. Rogers, B. Naydenov, F. Jelezko, S. Yamasaki, S. Nagamachi, T. Inubushi, N. Mizuochi, M. Hatano, *Sci. Rep.* **2015**, *5*, srep12882.

[15] L. J. Rogers, K. D. Jahnke, M. H. Metsch, A. Sipahigil, J. M. Binder, T. Teraji, H. Sumiya, J. Isoya, M. D. Lukin, P. Hemmer, F. Jelezko, *Phys. Rev. Lett.* **2014**, *113*, 263602.

[16] A. Sipahigil, R. E. Evans, D. D. Sukachev, M. J. Burek, J. Borregaard, M. K. Bhaskar, C. T. Nguyen, J. L. Pacheco, H. A. Atikian, C. Meuwly, R. M. Camacho, F. Jelezko, E. Bielejec, H. Park, M. Lončar, M. D. Lukin, *Science* **2016**, *354*, 847.

[17] J. R. Maze, P. L. Stanwix, J. S. Hodges, S. Hong, J. M. Taylor, P. Cappellaro, L. Jiang, M. V. G. Dutt, E. Togan, A. S. Zibrov, A. Yacoby, R. L. Walsworth, M. D. Lukin, *Nature* **2008**, *455*, 644.

[18] B. Hensen, H. Bernien, A. E. Dréau, A. Reiserer, N. Kalb, M. S. Blok, J. Ruitenberg, R. F. L. Vermeulen, R. N. Schouten, C. Abellán, W. Amaya, V. Pruneri, M. W. Mitchell, M. Markham, D. J. Twitchen, D. Elkouss, S. Wehner, T. H. Taminiau, R. Hanson, *Nature* **2015**, *526*, 682.

[19] F. Jelezko, T. Gaebel, I. Popa, M. Domhan, A. Gruber, J. Wrachtrup, *Phys. Rev. Lett.* **2004**, *93*, DOI 10.1103/PhysRevLett.93.130501.

[20] L. Childress, M. V. G. Dutt, J. M. Taylor, A. S. Zibrov, F. Jelezko, J. Wrachtrup, P. R. Hemmer, M. D. Lukin, *Science* **2006**, *314*, 281.

[21] M. V. G. Dutt, L. Childress, L. Jiang, E. Togan, J. Maze, F. Jelezko, A. S. Zibrov, P. R. Hemmer, M. D. Lukin, *Science* **2007**, *316*, 1312.

[22] G. Balasubramanian, I. Y. Chan, R. Kolesov, M. Al-Hmoud, J. Tisler, C. Shin, C. Kim, A. Wojcik, P. R. Hemmer, A. Krueger, T. Hanke, A. Leitenstorfer, R. Bratschitsch, F. Jelezko, J. Wrachtrup, *Nature* **2008**, *455*, 648.

[23] P. Maletinsky, S. Hong, M. S. Grinolds, B. Hausmann, M. D. Lukin, R. L. Walsworth, M. Loncar, A. Yacoby, *Nat. Nanotechnol.* **2012**, *7*, 320.

[24] M. S. Grinolds, S. Hong, P. Maletinsky, L. Luan, M. D. Lukin, R. L. Walsworth, A. Yacoby, *Nat. Phys.* **2013**, *9*, 215.

[25] M. S. Grinolds, M. Warner, K. De Greve, Y. Dovzhenko, L. Thiel, R. L. Walsworth, S. Hong, P. Maletinsky, A. Yacoby, *Nat. Nanotechnol.* **2014**, *9*, 279.

[26] J. M. Taylor, P. Cappellaro, L. Childress, L. Jiang, D. Budker, P. R. Hemmer, A. Yacoby, R. Walsworth, M. D. Lukin, *Nat. Phys.* **2008**, *4*, 810.

[27] T. M. Babinec, B. J. M. Hausmann, M. Khan, Y. Zhang, J. R. Maze, P. R. Hemmer, M. Lončar, *Nat. Nanotechnol.* **2010**, *5*, 195.

[28] M. J. Burek, Y. Chu, M. S. Z. Liddy, P. Patel, J. Rochman, S. Meesala, W. Hong, Q. Quan, M. D. Lukin, M. Lončar, *Nat. Commun.* **2014**, *5*, 5718.

[29] E. Neu, P. Appel, M. Ganzhorn, J. Miguel-Sánchez, M. Lesik, V. Mille, V. Jacques, A. Tallaire, J. Achard, P. Maletinsky, *Appl. Phys. Lett.* **2014**, *104*, 153108.

[30] J. P. Hadden, J. P. Harrison, A. C. Stanley-Clarke, L. Marseglia, Y.-L. D. Ho, B. R. Patton, J. L. O'Brien, J. G. Rarity, *Appl. Phys. Lett.* **2010**, *97*, 241901.

[31] S. A. Momenzadeh, R. J. Stöhr, F. F. de Oliveira, A. Brunner, A. Denisenko, S. Yang, F. Reinhard, J. Wrachtrup, *Nano Lett.* **2015**, *15*, 165.

[32] C. Du, T. van der Sar, T. X. Zhou, P. Upadhyaya, F. Casola, H. Zhang, M. C. Onbasli, C. A. Ross, R. L. Walsworth, Y. Tserkovnyak, A. Yacoby, *Science* **2017**, *357*, 195.

[33] P. Forsberg, M. Karlsson, *Opt. Express* **2013**, *21*, 2693.

[34] M. Karlsson, F. Nikolajeff, *Opt. Express* **2003**, *11*, 502.

[35] L. Li, T. Schröder, E. H. Chen, M. Walsh, I. Bayn, J. Goldstein, O. Gaathon, M. E. Trusheim, M. Lu, J. Mower, M. Cotlet, M. L. Markham, D. J. Twitchen, D. Englund, *Nat. Commun.* **2015**, *6*, 6173.

[36] B. Khanaliloo, M. Mitchell, A. C. Hryciw, P. E. Barclay, *Nano Lett.* **2015**, *15*, 5131.

[37] M. J. Burek, N. P. de Leon, B. J. Shields, B. J. M. Hausmann, Y. Chu, Q. Quan, A. S. Zibrov, H. Park, M. D. Lukin, M. Lončar, *Nano Lett.* **2012**, *12*, 6084.

[38] B. Khanaliloo, H. Jayakumar, A. C. Hryciw, D. P. Lake, H. Kaviani, P. E. Barclay, *Phys. Rev. X* **2015**, *5*, DOI 10.1103/PhysRevX.5.041051.

[39] M. J. Burek, D. Ramos, P. Patel, I. W. Frank, M. Lončar, *Appl. Phys. Lett.* **2013**, *103*, 131904.

[40] I. Bayn, S. Mouradian, L. Li, J. A. Goldstein, T. Schröder, J. Zheng, E. H. Chen, O. Gaathon, M. Lu, A. Stein, C. A. Ruggiero, J. Salzman, R. Kalish, D. Englund, *Appl. Phys. Lett.* **2014**, *105*, 211101.

[41] S. Mouradian, N. H. Wan, T. Schröder, D. Englund, *Appl. Phys. Lett.* **2017**, *111*, 021103.

[42] R. H. Telling, C. J. Pickard, M. C. Payne, J. E. Field, *Phys. Rev. Lett.* **2000**, *84*, 5160.

[43] F. K. De Theije, O. Roy, N. J. Van der Laag, W. J. P. Van Enckevort, *Diam. Relat. Mater.* **2000**, *9*, 929.

[44] S. Mouradian, N. H. Wan, T. Schröder, D. Englund, *ArXiv170407918 Cond-Mat Physicsphysics Physicsquant-Ph* **2017**.

[45] B. Chapman, *Glow Discharge Processes: Sputtering and Plasma Etching*, Wiley-Interscience, New York, **1980**.

[46] P. Appel, E. Neu, M. Ganzhorn, A. Barfuss, M. Batzer, M. Gratz, A. Tschöpe, P. Maletinsky, *Rev. Sci. Instrum.* **2016**, *87*, 063703.

[47] R. J. Epstein, F. M. Mendoza, Y. K. Kato, D. D. Awschalom, *Nat. Phys.* **2005**, *1*, 94.

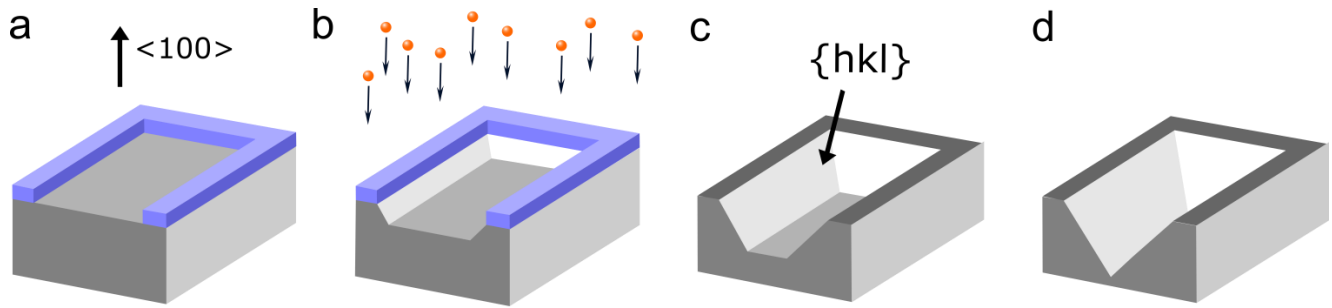

**Figure 1**. Schematics of crystal direction dependent etch in a monocrystalline material: a) An etch mask (blue) is lithographically patterned on (100) substrate surface. b) Reactive ions are accelerated by self-bias to bombard substrate surface, if etch rate in <100> is greater than that in <hkl>, tapered {hkl} planes will develop. c) After etching, the mask is removed and the etch profile is analogous to Si etch in KOH. d) For a longer etch time, crystal planes intercept at bottom.

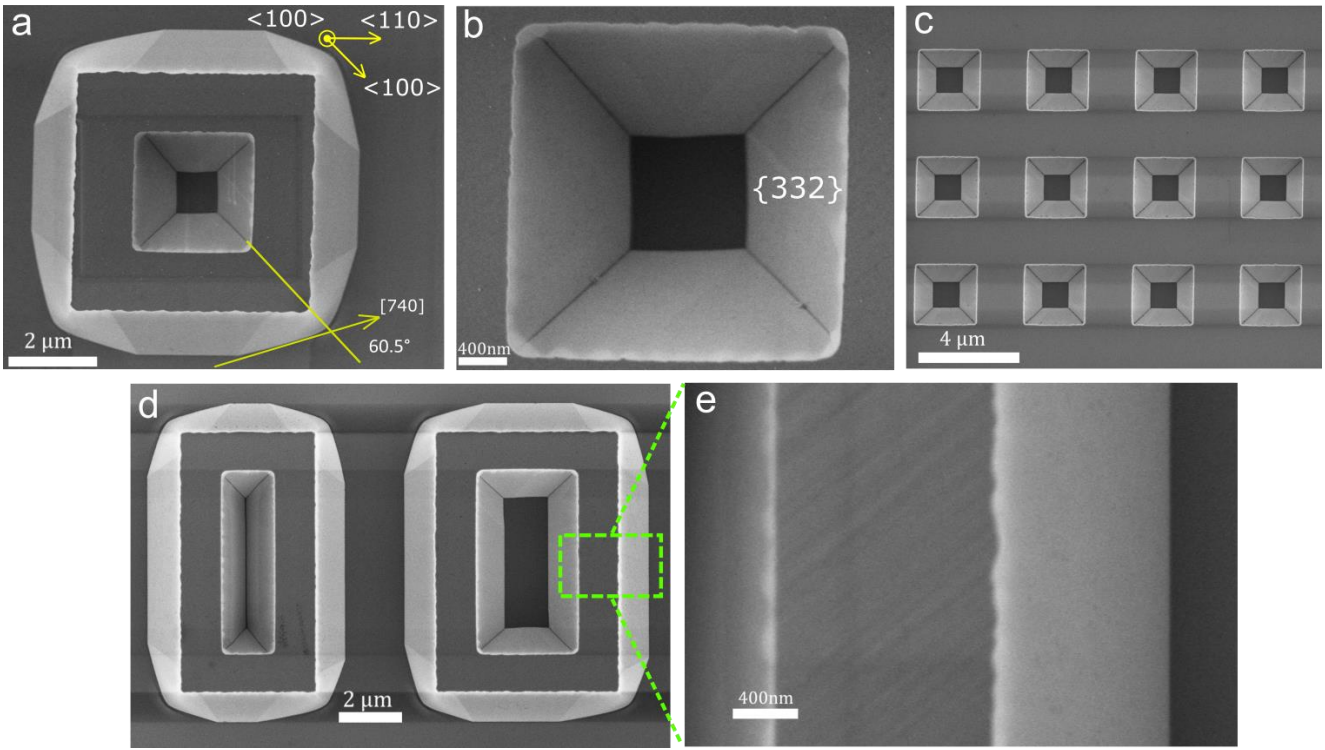

**Figure 2.** SEM images of the forms etched at 5 W substrate power for 70 min with window edges aligned parallel to <110>. a) Truncated square pyramid, extra facets appeared at outside corners. b) Zoom of a, tapered sidewalls belong to {332} family, sharp intersection lines at corners. c) An array of square features. d) Rectangular windows with different widths, the narrower one formed V-shaped groove. Slight horizontal stripes are caused by charging effect during SEM. e) A close-up of sidewall of etch profile in d, the fine intersection line between {332} sidewall and {100} bottom surface is shown.

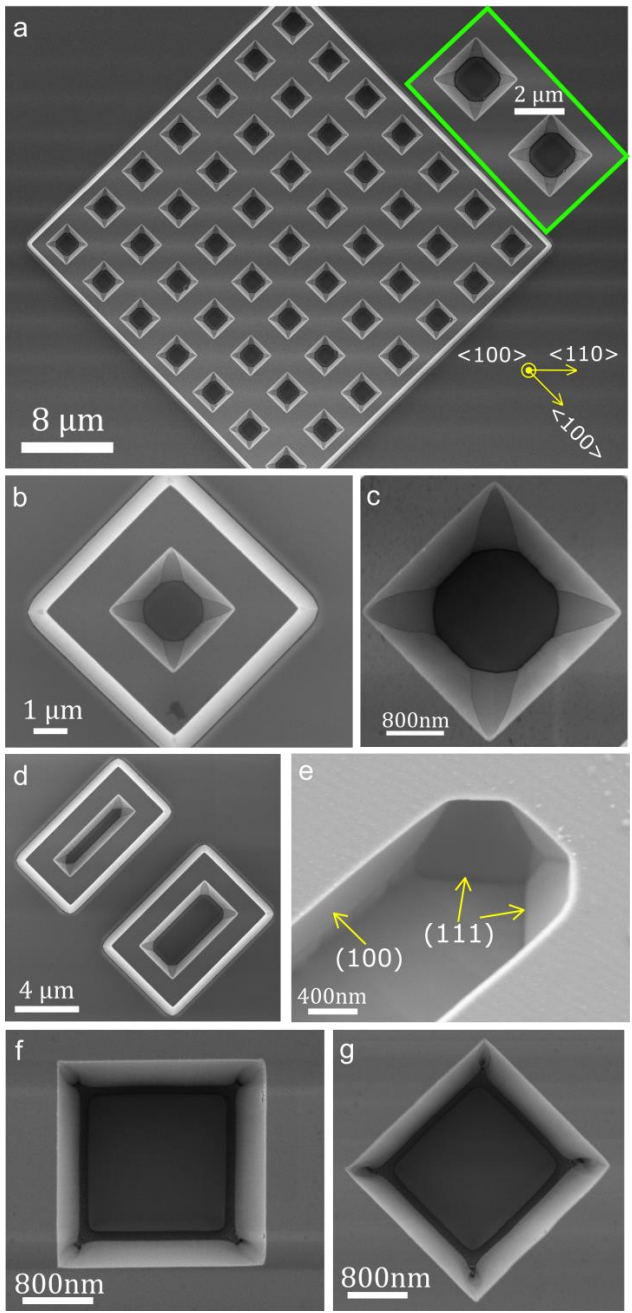

**Figure 3. a – d** SEM images of the forms etched at 5 W substrate power for 70 min with window edges parallel to <100>. a) An array of square patterns (close up in inset). b) An isolated square feature, no extra out-corner facets. c) Zoom in of b, newly emerged corner surfaces are close to {111}, their intersections with sidewalls form arc lines. d) Rectangular windows with different widths, due to smaller tapered angles of sidewalls, sidewalls do not merge at bottom. e) Appearance of facets {111}

at zero substrate power. f) & g) Disappearance of crystal anisotropy etch at 80W substrate power, identical etch morphology obtained in both <110> and <100> aligned windows.

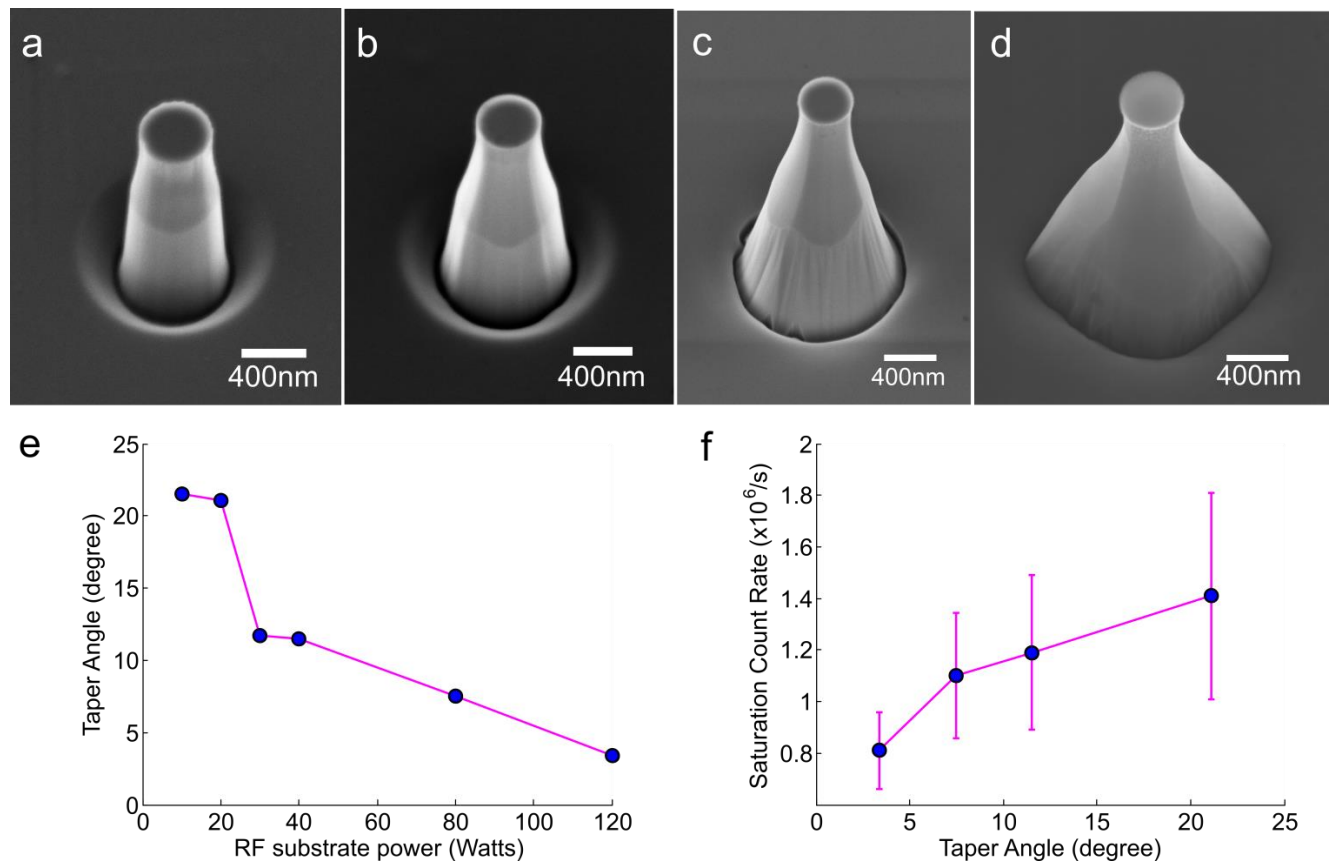

**Figure 4**. SEM images of monolithic diamond nanopillars showing tapering angles (half apex angle). a) 3.4°. b) 7.5°. c) 11.5°. d) 21°. e) Taper angle vs. RF substrate power, linear relationship when power ≥ 40 W and discrete relationship at lower power region. f) Saturation fluorescence count rate from a single NV center inside the nanopillar as function of tapering angle. The error bars correspond to the standard deviation of 10-20 nanopillars in one angle.